\documentclass[8.5pt,twoside,twocolumn]{article}
\usepackage[super,sort&compress,comma]{natbib}
\usepackage[version=3]{mhchem}
\usepackage{balance}
\usepackage{times,mathptm}
\usepackage{sectsty}
\usepackage{graphicx}
\usepackage{lastpage}
\usepackage[format=plain,justification=raggedright,singlelinecheck=false,font=small,labelfont=bf,labelsep=space]{caption}
\usepackage{color}

\begin{document}

\twocolumn[
  \begin{@twocolumnfalse}
\noindent\LARGE{\textbf{Fully spin-polarized quadratic non-Dirac bands realized quantum anomalous Hall effect$^\dag$}}
\vspace{0.6cm}

\noindent\large{\textbf{Ping Li Tian-Yi Cai$^{\ast}$\textit{$^{a}$}}}\vspace{0.5cm}

\noindent{\textbf{School of Physical Science and Technology, Soochow University, Suzhou 215006, People's Republic of China}}\vspace{0.5cm}

\end{@twocolumnfalse} \vspace{0.6cm}
  ]

\noindent\textbf{The quantum anomalous Hall effect is a intriguing quantum state which exhibits the chiral edge states in the absence of magnetic field. While the search for quantum anomalous Hall insulators is still active, the researchers mainly search for the systems containing magnetic atom. Here, based on first-principles density functional theory, we predict a new family of Chern insulators with fully spin-polarized quadratic \emph{p$_{x,y}$} non-Dirac bands in the alkali earth metal BaX (X = Si, Ge, Sn) system. We show that BaX monolayer has a half-metallic ferromagnetic ground state. The ferromagnetism is mainly originated from the $\emph{p}$ orbitals of Si, Ge and Sn atoms. The 2D BaSn monolayer exhibits a large magnetocrystalline anisotropic energy of 12.20 meV/cell and a nontrivial band gap of 159.10 meV. Interestingly, both the chiral edge currents direction and the sign of Chern number can be tuned by doping. Furthermore, the 4$\%$ compressive strain in the 2D BaX systems can drive structural phase transition but the nontrivial topological properties remain reserved. Our findings not only extend the novel topological physics but also provide fascinating opportunities for the realization of quantum anomalous Hall effect experimentally.}

\section{Introduction}

Two-dimensional (2D) magnetic topological states have attracted tremendous attention due to their exotic physical phenomena and unique properties contrasting to their bulk counterparts. \cite{1,2,3,4,5,6} In particular, the quantum anomalous Hall (QAH) phase is characterized by a chiral edge state current that encloses an insulating bulk. The QAH effect requires both the broken time reversal symmetry and spin-orbit coupling (SOC). \cite{7,8,9} More interestingly, the QAH effect is featured by a quantized Hall conductance $\sigma_{xy}$ = Ce$^2$/h, where C is the Chern number, and amounts to the number of chiral edge states. \cite{10} The chiral edge states are topologically protected and robust against the electron scattering, which provides great potential applications for designing low energy consumption and dissipationless spintronic devices.

Until now, only the V or Cr-doped (Bi,Sb)$_2$Te$_3$ systems have experimentally shown a plateau in the Hall conductance with a range of the gate voltage. \cite{11,12,13} The experimental conditions are extreme requirements, such as the extremely low temperature ($<$100 mK) due to the small band gap and the greatly accurate controlling of the extrinsic impurities. It greatly hinder their further device applications. Hence, it would be meaningful to search for a new Chern insulator with large band gap and high Curie temperature ({$\rm T_C$}).

Recently, the enormous Chern insulators have been theoretically predicted. \cite{14,15,16,17,18,19,20,21} The QAH effect, which persists in the absence of an external magnetic field, can be realized in systems with the magnetic atom or the thin slab of three-dimensional (3D) topological insulator by magnetic dopants breaking the time reversal symmetry (TRS). These systems have a common feature, namely the existence of magnetic atoms. Simultaneously, the alkali earth metal BaX (X = Si, Ge, Sn) system is found to be a unique ferromagnetic (FM) half-metal.\cite{22,23,24,25} Interestingly, these compounds do not include any transition metal atoms. Hence, the mechanism of the ferromagnetism is different from the $\emph{p}$-$\emph{d}$ exchange and double exchange that are crucial in the conventional magnetic materials. The magnetism originates from the spin polarization of the $\emph{p}$ states of anions.

In this paper, based on the density functional theory, we report that the QAH effect can be realized in the BaX monolayer with fully spin-polarized quadratic $\emph{p$_{x,y}$}$ non-Dirac bands. It is well-known that the linear-dispersive Dirac bands are consisted of $\emph{p$_{z}$}$ orbital around the K point, as observed in graphene. However, the quadratic non-Dirac bands are composed of $\emph{p$_{x,y}$}$ orbitals at the $\Gamma$ point due to the trigonal symmetry of BaX monolayers. It is reported that a non-trivial SOC gap can be induced with a minimal basis of three orbitals ($\emph{s}$, $\emph{p$_x$}$, $\emph{p$_y$}$) in a 2D trigonal lattice.\cite{26} Different from the $\emph{p$_{z}$}$ orbitals linear Dirac topological states which is easily destroyed by the substrates, \cite{27,28} the $\emph{p$_{x,y}$}$ orbitals formed $\sigma$-bond is particularly robust. \cite{19,29,30} In addition, the absence of imaginary frequencies in the calculated phonon spectra confirms the dynamic stabilization of the monolayer BaX. Moreover, the band gap at the quadratic non-Dirac point can be opened by the SOC. The calculated quantized Hall conductance, Chern number, Berry curvature and edge states further indicate the nontrivial topology. More interestingly, we find that the nontrivial topology is robust against the biaxial strain with structural phase transitions. These findings provide broader opportunities for the investigation of the QAH effect at high temperatures.

\section{Computational methods}
To investigate the electronic and magnetic structures, we implemented the Vienna $Ab$ $initio$ Simulation Package (VASP) \cite{31,32} for the first-principles calculations based on density functional theory (DFT). The electron exchange-correlation functional was described by the generalized gradient approximation of the Perdew-Burke-Ernzerhof functional. \cite{33} The plane-wave basis set with a kinetic energy cutoff of 500 eV was employed. Here, $12\times 12\times 1$ and $24\times 24\times 1$ $\Gamma$-centered $k$ meshes are adopted for the structural optimization and the self-consistent calculations. To avoid unnecessary interactions between the monolayer, the vacuum layer was set to 20 \AA. The total energy convergence criterion was set to be 10$^{-6}$ eV. To confirm the structural stability, the phonon spectra were calculated using a finite displacement approach as implemented in the PHONOPY code, in which a $4\times 4\times 1$ supercell were used. \cite{34} An effective tight-binding Hamiltonian constructed from the maximally localized Wannier functions (MLWFs) was employed to explored the edge states. \cite{35,36,37} Therefore, the edge states were calculated in a half-infinite boundary condition using the iterative Green's function method by the package WANNIERTOOLS. \cite{37,38}

\section{Results and discussion}

\subsection{Structure and stability}
A possible BaX monolayer structure is shown in Figure 1 (a,b). The primitive unit cell consists of one Ba atom and one X (Si, Ge, Sn) atom forming a hexagonal honeycomb lattice. The material is entirely flat in a single atomic layer as graphene. Hence, the space group is $\emph{P-6m2}$, and the point group is $\emph{D$_{3h}$}$. The optimized lattice constant a$_m$ of BaX monolayer are listed in Table I.

\begin{table}[htbp]
\caption{
The lattice constants a$_m$ (a$_b$) [(\AA)] for the monolayer (bulk), cohesive energy E$_c$ (eV/atom), formation energy E$_f$ (eV/atom), magnetocrystalline anisotropy energy (MAE) [meV/cell],  global band gap E$_g$ (meV) and Curie temperature {$\rm T_C$} (K) of the alkali earth metal BaX (X = Si, Ge, Sn).}
\begin{tabular}{cccccccc}
  \hline
         & a$_m$  & a$_b$   & E$_c$   & E$_f$  & MAE     & E$_g$     & {$\rm T_C$}    \\
  \hline
BaSi     & 5.57    & 4.02   & 2.53    & 0.93   & -1.17   & 2.70       & 284   \\
BaGe     & 5.61    & 4.07   & 2.41    & 0.94   &  1.06   & 47.21      & 286   \\
BaSn     & 5.95    & 4.26   & 2.19    & 1.03   & 12.20   & 159.10     & 356   \\
  \hline
\end{tabular}
\end{table}

The structural stability of BaX monolayer is verified by the following important aspects: (1) cohesive energy, (2) formation energy, and (3) dynamical stability. Firstly, we calculate the cohesive energy E$_c$, defined as E$_c$ = ($\mu_{Ba}$ + $\mu_{X}$ -E$_{tot}$)/N. Here, $\mu_{Ba}$ and $\mu_{X}$ are the chemical potential of Ba and X atom which are chosen to be the total energy of an isolated Ba and X atom, respectively. E$_{tot}$ is the total energy of BaX monolayer. N is the number of atoms in the primitive unit cell. The calculated E$_c$ are illustrated in Table I. The cohesive energies are about 2.19 - 2.53 eV/atom larger than that of the bismuthene (1.954 eV/atom). \cite{39} Since bismuthene have recently been experimentally synthesized, \cite{29,40} it indicates that BaX monolayer could be synthesized experimentally. The formation energy of BaX monolayer is defined as E$_f$ = E$_{2D}$/N$_{2D}$ - E$_{3D}$/N$_{3D}$, \cite{41,42} where E$_{2D}$ and E$_{3D}$ are the total energies of 2D and 3D structure, respectively. N$_{2D}$ and N$_{3D}$ denote the number of atoms in the respective unit cells. The formation energies are about 0.93 - 1.03 eV/atom, which are larger than that of silicene (0.76 eV/atom). \cite{43,44,45,46} Finally, the dynamical stability can be reflected in the phonon spectrum. As illustrated in Figure 1 (c-d), all the vibrational modes show positive frequencies, indicating the stabilization of the structure shown in Figure 1(a).

\begin{figure}[htb]
\begin{center}
\includegraphics[angle=0,width=1.00\linewidth]{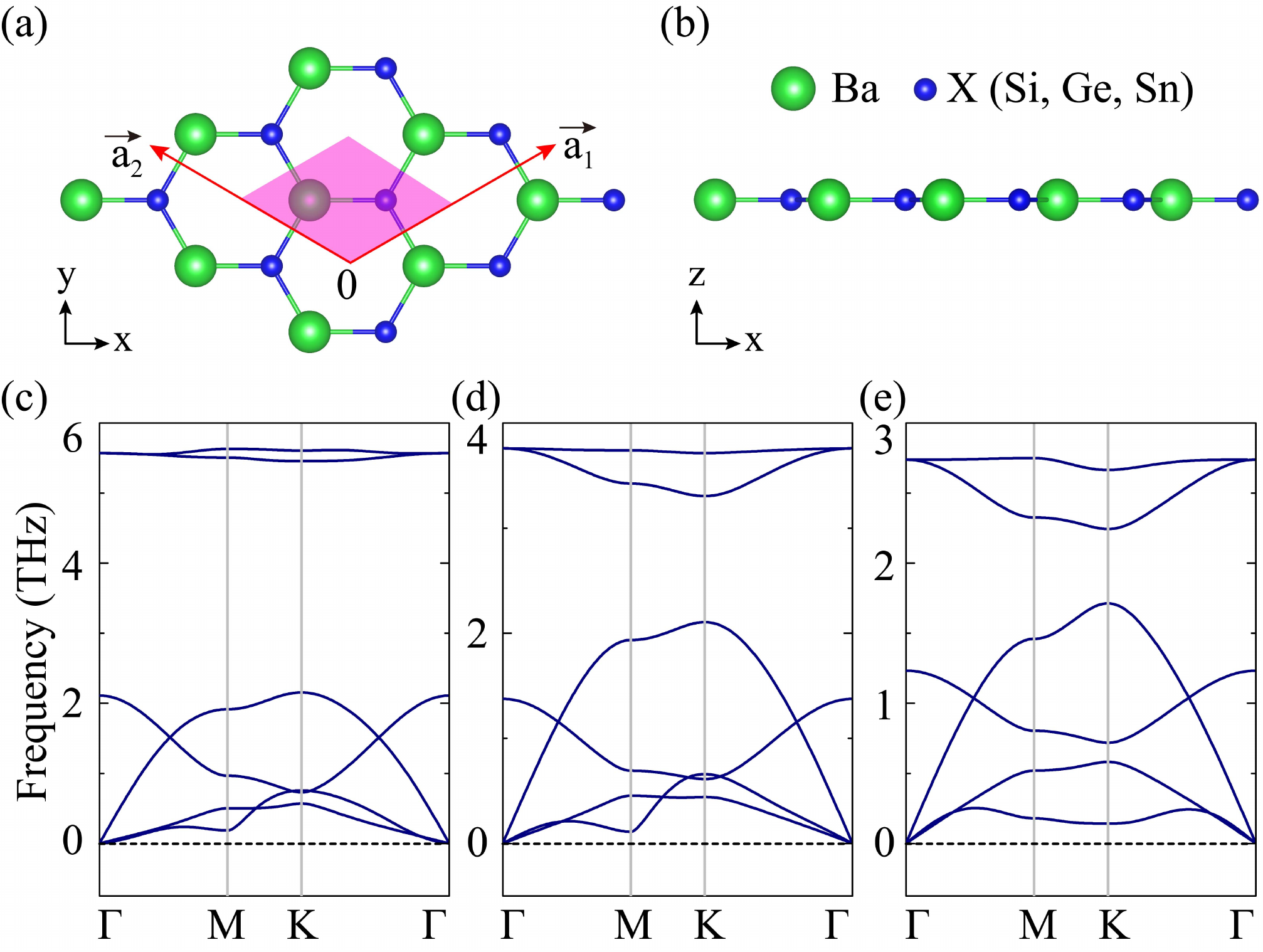}
\caption{(a) Top and (b) side view of the lattice structure for BaX (X = Si, Ge, Sn) monolayer with lattice vectors $\vec{a}_1$ and $\vec{a}_2$, the unit cell is indicated by the magenta shading. The Ba and X (X = Si, Ge, Sn) atoms are depicted by the green and blue balls, respectively. (c-e) The calculated phonon dispersion curves of (c) BaSi, (d) BaGe and (e) BaSn.}
\end{center}
\end{figure}

In order to further check whether the BaX has a buckled structure, we construct the buckled BaX structure, similar to the case of silicene, germanene and stanene. The calculated energy differences between the buckled and planar structure are illustrated in Figure 2. Remarkably, the energies of BaX monolayer monotonically increase with the increase of buckled height (d) from 0 to 1.4 \AA. It shows that the planar structure has the lowest energy. Therefore, the planar structure is the most stable structure.

\begin{figure}[htb]
\begin{center}
\includegraphics[angle=0,width=1.00\linewidth]{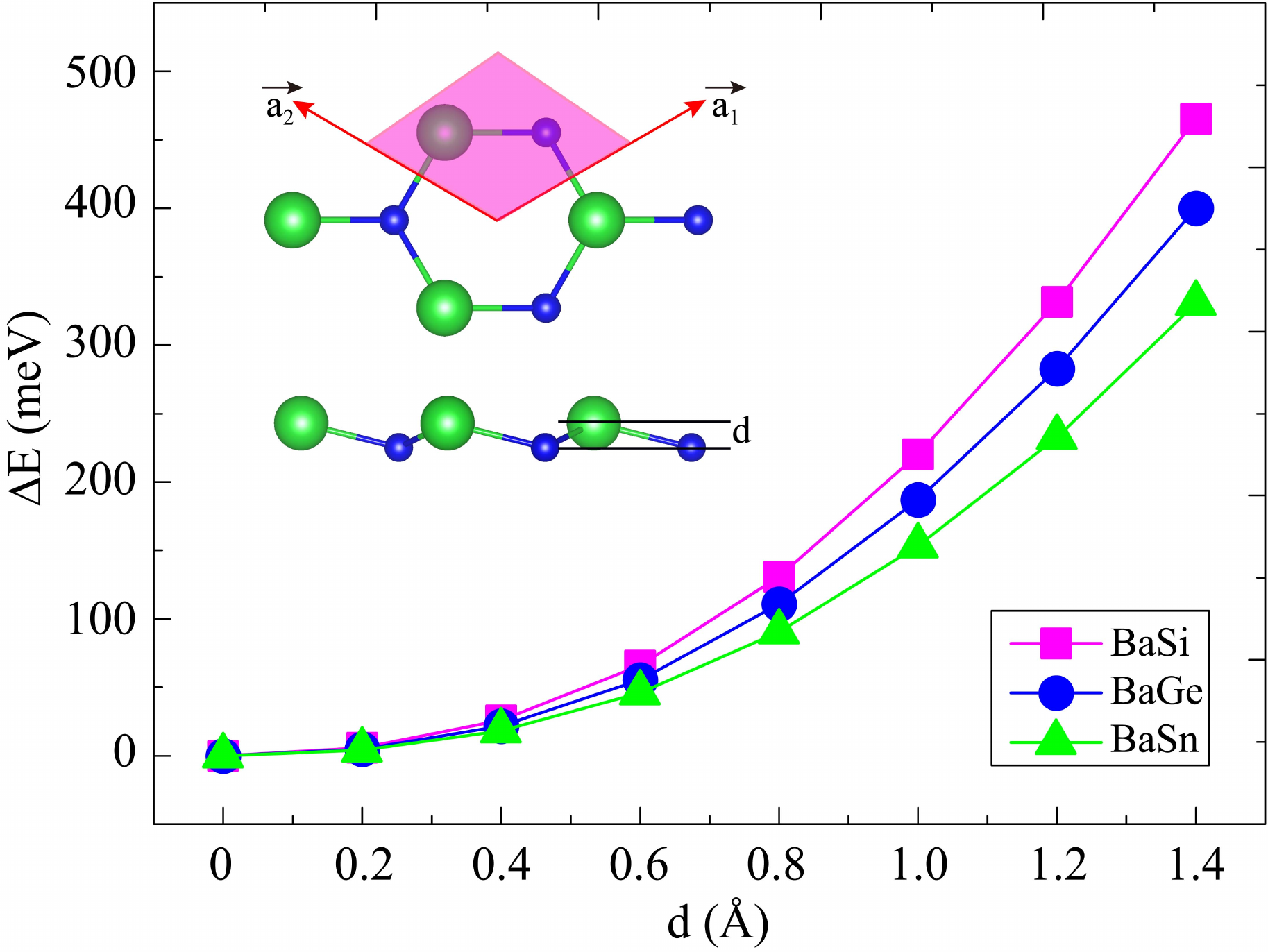}
\caption{ The variation of total energy difference, $\Delta E$ = E(buckled) - E(planar), as a function of buckled height d. The insets is the buckled structure of BaX monolayer. }
\end{center}
\end{figure}

\subsection{Magnetic property}
Our calculations show that the total energy of the non-spin-polarized BaX monolayer is higher than that of the spin-polarized BaX monolayer. The energy difference are 0.13 eV/atom (BaSi), 0.12 eV/atom (BaGe) and 0.13 eV/atom (BaSn), respectively. To understand the magnetic interactions, we consider all possible magnetic configurations in the 2$\times$2 supercell, namely, nonmagnetic (NM), FM, and antiferromagnetic (AFM). We find that the FM state is the ground state. The AFM state is 156.83 meV, 158.08 meV and 197.22 meV higher than the FM state for BaSi, BaGe and BaSn, respectively. The FM ground state has an integer magnetic moment of 2 $\mu_B$ per unit cell. The magnetic moment is mainly originated from the $\emph{p}$ orbitals of the X atom, while the magnetic moment on the Ba atom can be ignored. It is consistent with their bulk structures. \cite{22,23,24,25}

In addition, we calculate the magnetocrystalline anisotropy energies (MAE) of the BaX by considering the SOC effect. Here, two magnetization easy-axis directions, [100] (in plane) and [001] (out of plane), are considered. The MAE is defined as $\Delta E$ = E$_{100}$ - E$_{001}$. The positive value of MAE shows the easy axis is along the z axis rather than the x axis. The calculated MAE are listed in Table I. We find that only the magnetization direction of BaSi is along the x axis. The large MAE value of BaSn reaches 12.20 meV/cell, which is nine times larger than that of 2D CrI$_3$ (1.37 meV/cell). \cite{47} Experimentally, it is reported that the magnetization direction of 2D ferromagnets could be tuned by an external field in Fe$_3$GeTe$_2$ system. \cite{48} In our work, we will concentrate on the topological properties of BaX with the magnetization direction along the z axis. And the results with the magnetization direction along the x axis are presented in the supporting materials. As a result, our calculations show that the band topology is related to the direction of the easy axis.

The Curie temperature is important for the experimental observation of the QAH effect. Based on the Ising model, we estimate the Curie temperature by using Monte Carlo (MC) simulations. The nearest-neighbor exchange interaction parameters $\emph{J}$ can be obtained from the Hamiltonian of the Ising model:
\begin{equation}
H= -\sum_{i,j}JS_i\cdot S_j,
\end{equation}
where $\emph{S}$ is the magnetic moment per X atom, $\emph{i}$ and $\emph{j}$ stand for the nearest site pairs. The exchange interaction parameter $\emph{J}$ is related to the exchange energy, $\emph{J}$ = E$_{ex}$/12$\emph{S}$$^2$, where E$_{ex}$ = (E$_{AFM}$ - E$_{FM}$) is the energy difference between the antiferromagnetic and ferromagnetic state. Therefore, $\emph{J}$ can be calculated to be 13.07 meV for BaSi, 13.17 meV for BaGe and 16.44 meV for BaSn, respectively. Here, the MC simulations are implemented on a 80$\times$80 supercell which is adopted to reduce translational constraint, using 1$\times$10$^7$ loops for each temperature. The calculated Curie temperature are 284 K for BaSi, 286 K for BaGe and 356 K for BaSn, respectively.

\subsection{Electronic band structure}

\begin{figure}[htb]
\begin{center}
\includegraphics[angle=0,width=1.00\linewidth]{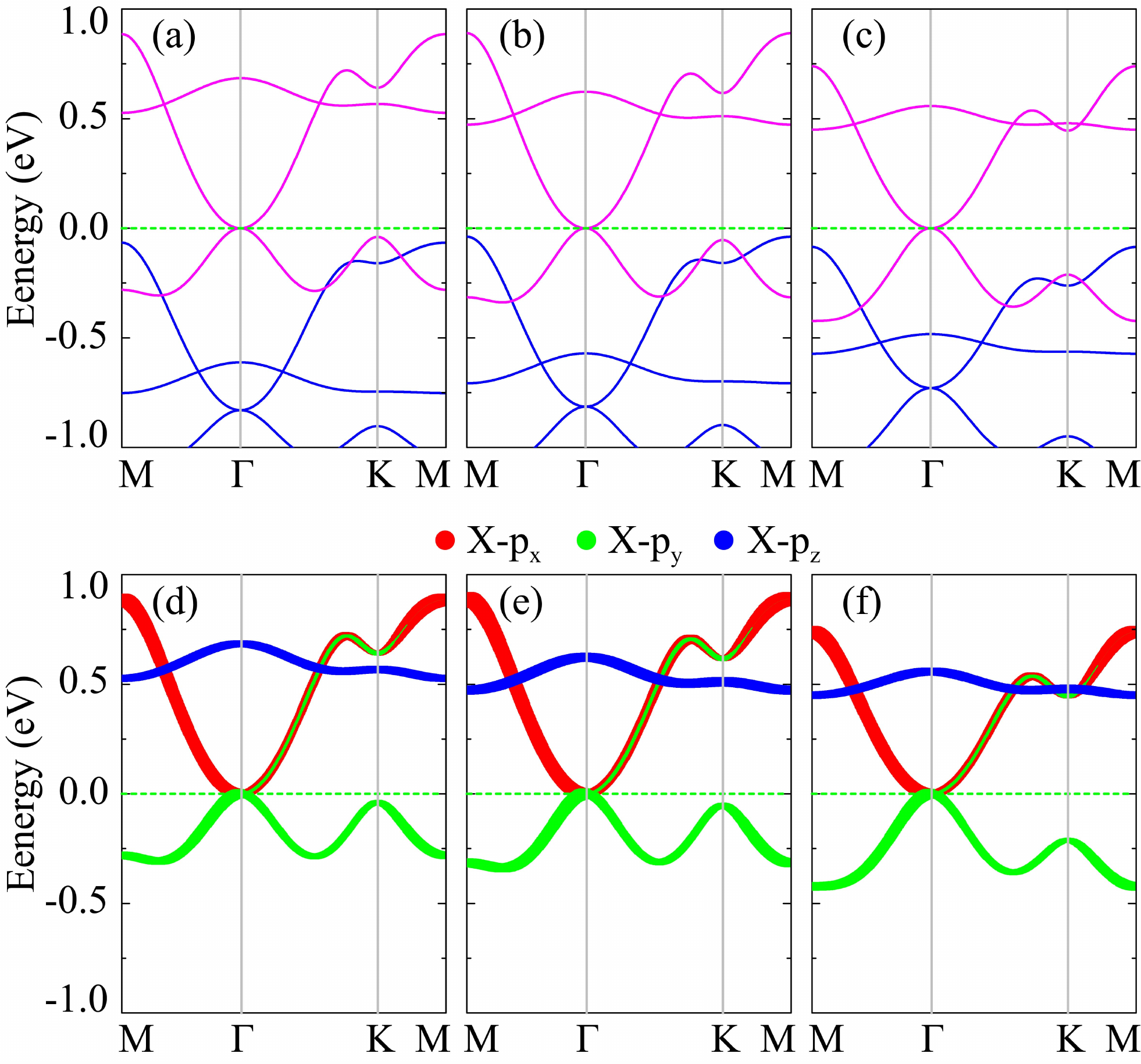}
\caption{(a-c) Spin-polarized band structures for (a) BaSi, (b) BaGe and (c) BaSn. The blue and magenta curves correspond to the majority and minority spin bands structure, respectively. (d-f) Energy and k contribution of X-p-resolved to the minority spin bands for (d) BaSi, (e) BaGe and (f) BaSn.  }
\end{center}
\end{figure}

The most fascinating property of monolayer BaX is shown in its electronic band structure. Firstly, the band structure without considering SOC effect is shown in Figure 3(a-c). Remarkably, the BaX monolayer is fully spin-polarized. The majority spin band is insulating while the minority spin band is metallic. One can see that the minority spin bands exhibit quadratic non-Dirac dispersions near the Fermi energy. In order to know its origin, the orbital resolved minority spin band structures are calculated, as shown in Figure 3(d-f). We find that the quadratic non-Dirac bands are mainly composed of X-$\emph{p$_{x,y}$}$ orbitals. As shown in Figure S3, the charge density of the non-Dirac state are calculated for BaSi, BaGe and BaSn, respectively. It can be clearly seen that the bond in plane has the nature of $\sigma$ orbital. It is noted that the $\emph{p$_{x,y}$}$ quadratic non-Dirac states are particularly robust because of the nature of $\sigma$ bonds, \cite{49} which is different from the $\emph{p$_{z}$}$ Dirac states due to the nature of weak $\pi$ bonds.\cite{28,50}

\begin{figure}[htb]
\begin{center}
\includegraphics[angle=0,width=0.80\linewidth]{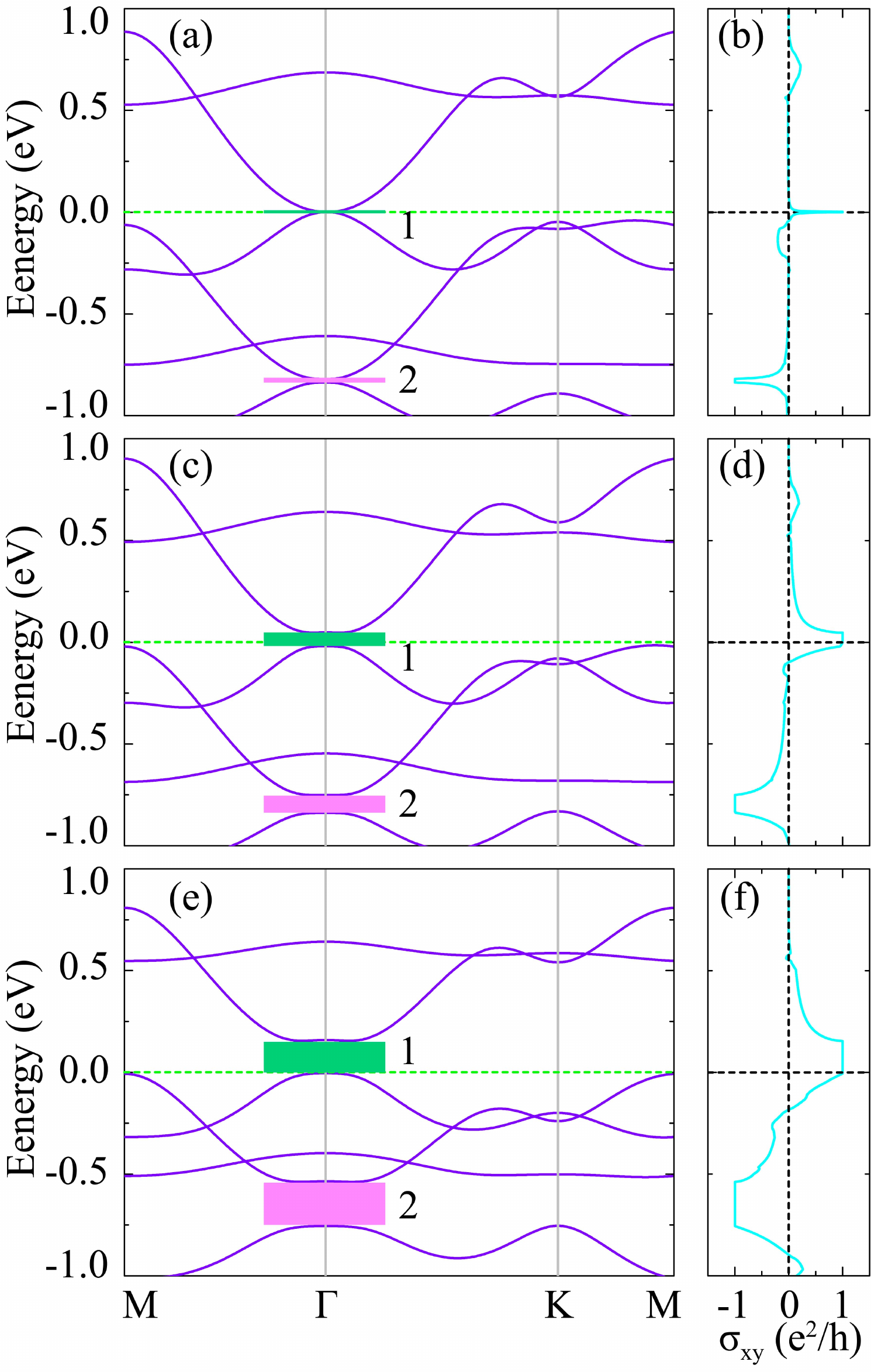}
\caption{ Band structures with SOC as well as anomalous Hall conductvity ($\sigma_{xy}$) of the (a, b) BaSi, (c, d) BaGe and (e, f ) BaSn with [001] magnetization direction. In the left panels, the band gap between minority (majority) spin bands is labeled as gap 1 (2).}
\end{center}
\end{figure}

Then, considering the SOC effect, Figure 4(a,c,e) shows the band structure of BaX monolayer. The indirect global band gap is opened by the SOC effect with the magnitude of 2.70 - 159.10 meV. Even for BaSi with the lightest Si atom, the quadratic non-Dirac dispersion near $E_F$ is still preserved. The large SOC of BaSn induces a gap of 163.19 meV at the high symmetry $\Gamma$ point. The large band gap indicates the application potential of BaSn in the spintronic devices.

\subsection{Quantum anomalous Hall effect}
To show the topological properties of BaX monolayer, we calculate the anomalous Hall conductivity (AHC) by using the formula£º
\begin{equation}
\sigma_{xy} = C\frac{e^2}{h},
\end{equation}
\begin{equation}
C= \frac{1}{2\pi} \int_{BZ} d^2k ~\Omega(\textbf{k}),
\end{equation}
where C is related to the quantized anomalous Hall conductance $\sigma_{xy}$, namely, Chern number. \cite{51} The right panel of Figure 4 shows the variation of $\sigma_{xy}$ with the Fermi energy. Obviously, two quantized steps in $\sigma_{xy}$ curves are observed, which is the characteristic of Chern insulator. $\sigma_{xy}$ has a quantized value (C = 1) when $E_F$ is in gap 1. When $E_F$ shifts down into the gap 2, Chern number will change to -1. It implies that the hole doping with the concentration of 3e per unit cell will reverse the direction of the chiral edge current. After doping, the ground state is still FM state. The calculated energy difference of AFM and FM states are listed in Table SI. The energies of AFM states are always higher than ones of FM states. In addition, the MAE in all cases are positive, which indicate the easy magnetization direction of the z axis.

\begin{figure}[htb]
\begin{center}
\includegraphics[angle=0,width=1.00\linewidth]{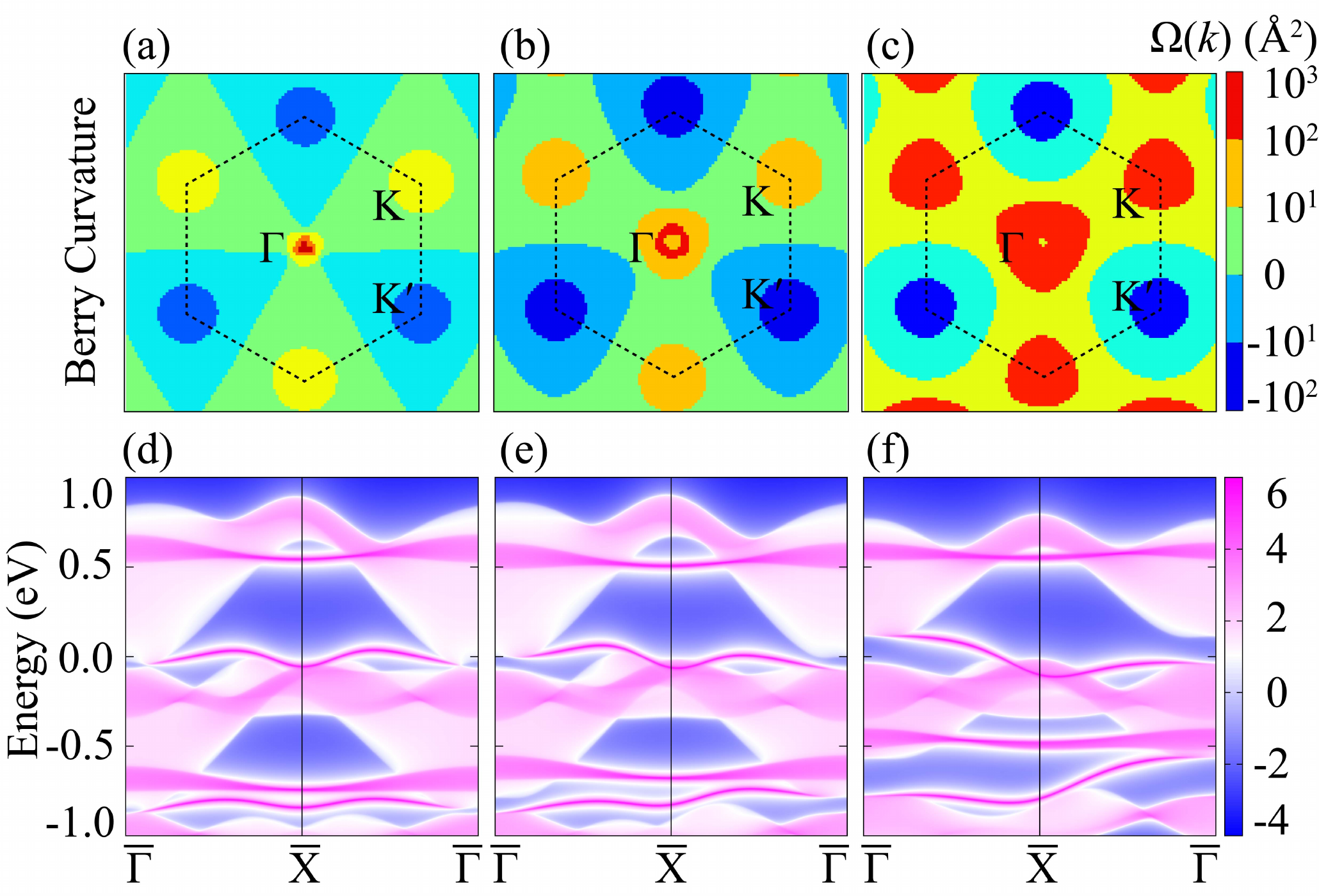}
\caption{(a-c) The Berry curvature with SOC in the momentum space and (d-f) calculated edge state of a semi-infinite sheet of the (a, d) BaSi, (b, e) BaGe and (c, f) BaSn with [001] magnetization direction. }
\end{center}
\end{figure}

In addition,
\begin{equation}
\Omega(\textbf{k})=-\sum_{n}f_{n}\sum_{n\prime \neq n}\frac{2Im \left \langle \psi_{nk} \mid v_{x} \mid \psi_{n\prime k} \right \rangle \left \langle \psi_{n\prime k} \mid v_{y} \mid \psi_{nk} \right \rangle}{(E_{n\prime}-E_{n})^2},
\end{equation}
where $\Omega(\textbf{k})$ is the Berry curvature in the reciprocal space, $v_{x}$ and $v_{y}$ are operator components along the x and y directions and $f_{n}=1$ for the occupied bands. \cite{52} The Berry curvature in the Brillouin zone are displayed in Figure 5(a-c). The Berry curvature shows the maximum values around the $\Gamma$ point, which contributes to the QAH effect.

Another prominent characteristic of Chern insulator is the existence of chiral edge states. We construct the Green's function of the BaX semi-infinite sheet from the MLWFs and calculate the local density of states at the edge, as shown in Figure 5(d-f). \cite{53,54} Within the gap 1 and 2, we can clearly see that one chiral edge state connects the valence and conduction bands. The number of edge states indicates the absolute value of the Chern number, which is $|C|$ = 1. \cite{55} Our findings suggest the realization of the QAH state and the direction control of the edge current.

\subsection{Strain induced structural phase transitions}

\begin{figure}[htb]
\begin{center}
\includegraphics[angle=0,width=0.90\linewidth]{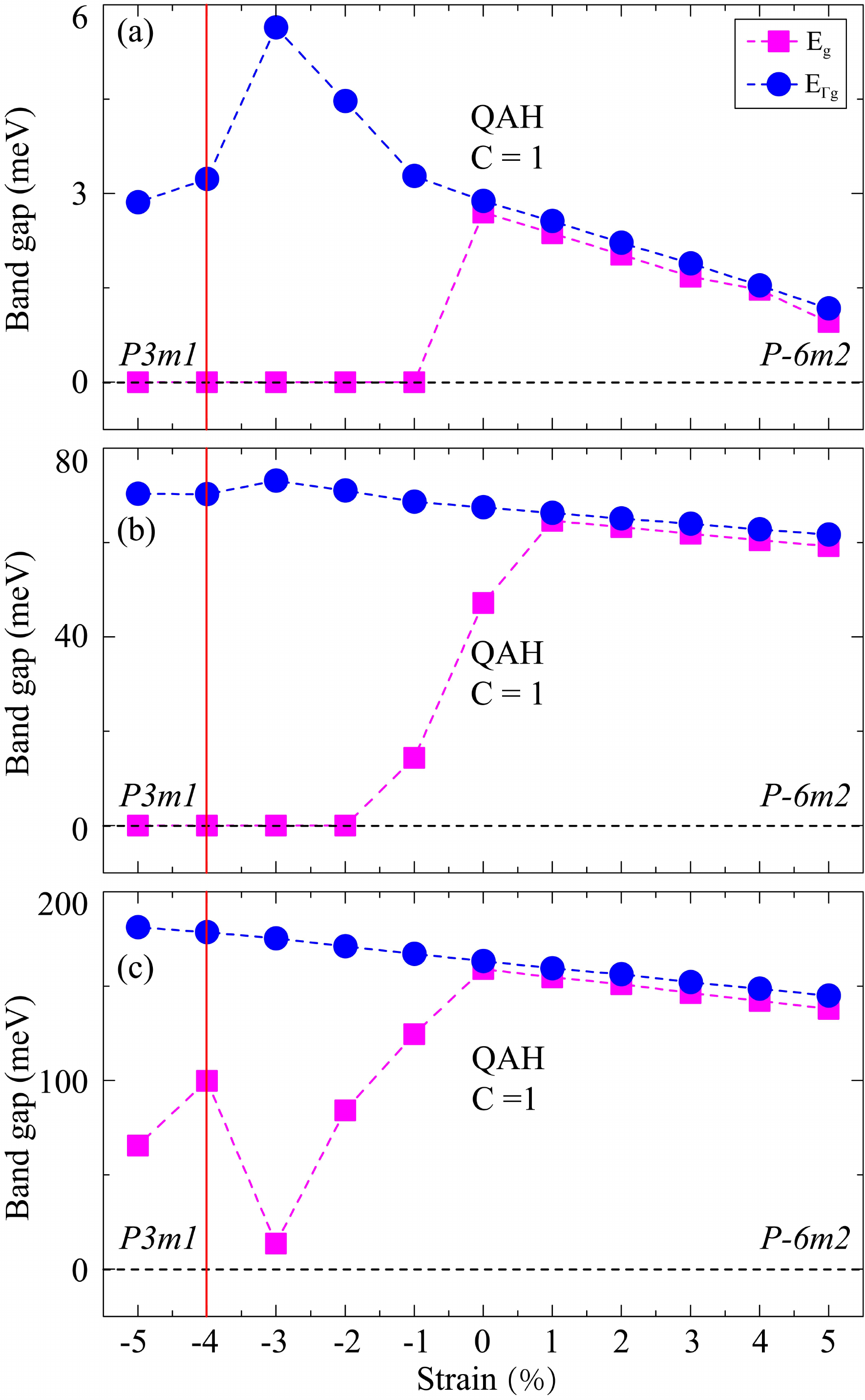}
\caption{ Calculated global gap (E$_g$) and high symmetry $\Gamma$ point gap (E$_{\Gamma g}$) as a function of the biaxial strain for (a) BaSi, (b) BaGe and (c) BaSn. The red solid line denotes the structural phase boundaries.}
\end{center}
\end{figure}

In 2D materials, strain effect can modulate the structure, magnetism, electronic structure and topological property. \cite{56,57,58} Hence, it is meaningful to explore these effects in BaX monolayer. In the previous studies, it was reported that the trivial semiconductor GaS and GaSe can be driven into a topologically nontrivial state by the biaxial strain. \cite{59} Thus, one immense concern is to examine the strain tolerance of topological properties in these three systems. In this paper, the in-plane biaxial strain is applied ranging from -5$\%$ to 5$\%$, which is defined as $\varepsilon$ = (a-a$_0$)/a$_0$$\times$100$\%$. Here a and a$_0$ represent the in-plane lattice constant after and before the strain is applied, respectively.

As shown in Figure 6, the calculated band gap is varied with the strain. 4$\%$ compressive strain can drive a structural phase transition from the planar structure ($\emph{P-6m2}$) to the buckled structure ($\emph{P3m1}$). For BaSi, BaGe and BaSn the topological properties will not be destroyed even with 5$\%$ extensive strain. However, only a small compressive strain can drive BaSi and BaGe transit into the metallic states. Importantly, BaSn monolayer maintain nontrivial topological phase within a relatively wide strain range of $\pm$5$\%$. Obviously, the robustness of topology against lattice deformation makes it easier for the experimental realization.

\subsection{The experimental synthesis}
The stabilities of 2D materials are important for the experimental fabrication and practical applications. Considering their good cohesive energy, formation energy and dynamic stabilities, we propose two methods to synthesize these 2D BaX monolayer in experiment. On one hand, BaX monolayer can be synthesized by a spontaneous conversion from a (111)-oriented cubic BaX to a layered structure,\cite{60} as shown in Figure 7(a). On the other hand, the molecular beam epitaxy (MBE) can fabricate BaX monolayer. In experiments, it has been testified that 2D CuSe monolayer with similar structure can be fabricated via the MBE approach. \cite{61} Figure 7(b) is a schematic diagram that the 2D BaX monolayer can be synthesized by the MBE method on the SiC(0001) substrate. The lattice mismatch is defined as $|$a$_{SiC}$ - a$_{BaX}$$|$/a$_{SiC}$, where a$_{SiC}$ and a$_{BaX}$ are the lattice constants of the SiC(0001) substrate and BaX monolayer, respectively. We noticed that the $1\times 1$ unit cell of BaX is commensurate to the $\surd3\times \surd3$/$2\times 2$ surface unit cell of SiC(0001) substrate. The BaX lattice constant are listed in Table I, and SiC(0001) lattice constant is 3.08 \AA. The lattice mismatch of less than 5$\%$ is then calculated. These effective methods can be generalized to synthesize other 2D materials.

\begin{figure}[htb]
\begin{center}
\includegraphics[angle=0,width=1.00\linewidth]{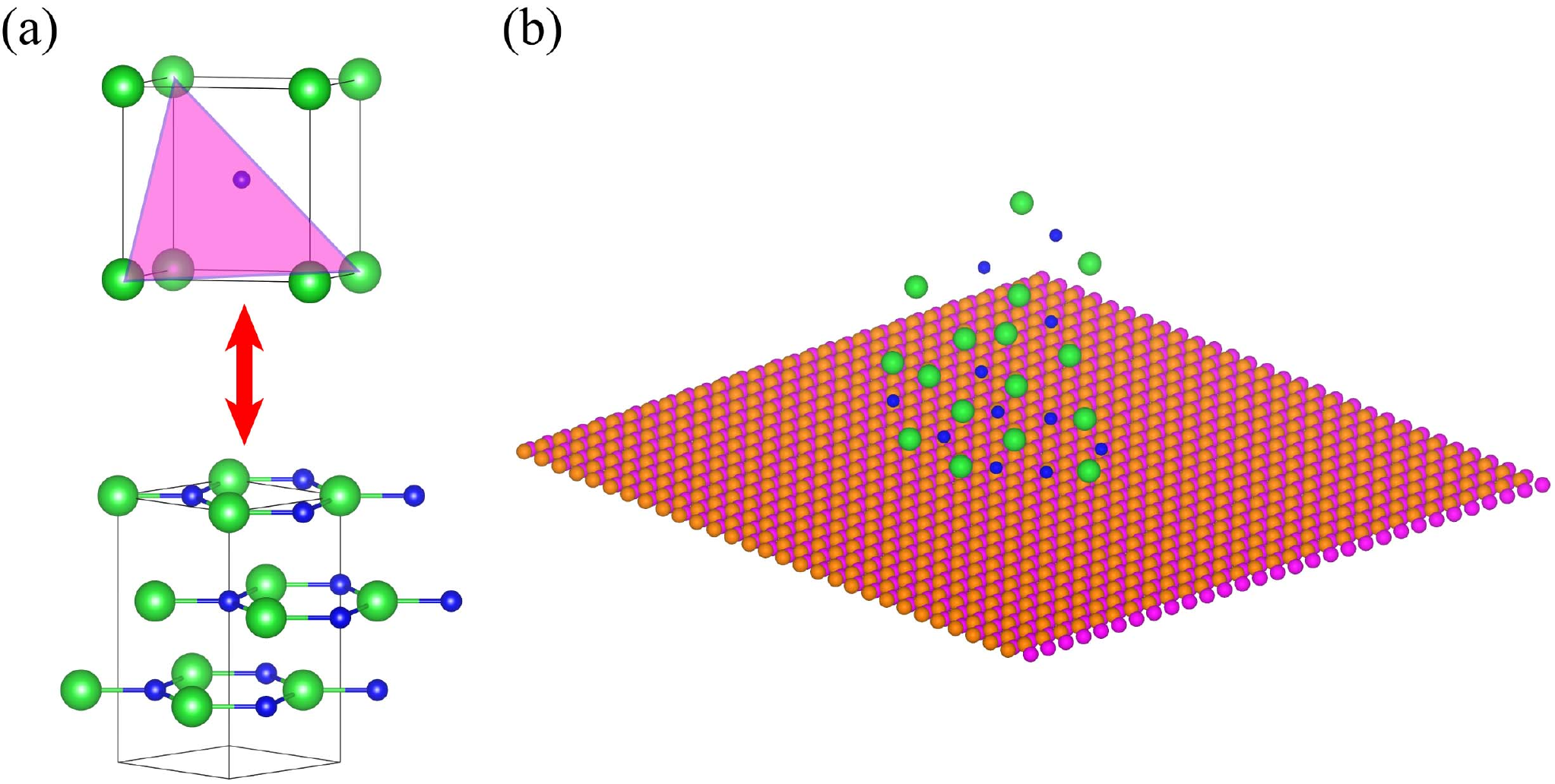}
\caption{Schematic describing the synthesis of the BaX (X = Si, Ge, Sn) monolayer. (a) The crystal structure of bulk BaX (X = Si, Ge, Sn). The (111) plane is highlighted by magenta color. The color codes for atoms are the same as the Figure 1(a, b). (b) It is a schematic diagram that the BaX (X = Si, Ge, Sn) monolayer can be synthesized on SiC(0001) substrate via the molecular beam epitaxy (MBE) approach. }
\end{center}
\end{figure}

\section{Conclusion}
In summary, using first-principles calculations, we predict a new class of Chern insulators candidates, namely ferromagnetic BaX (X = Si, Ge, Sn) monolayer. The quadratic $\emph{p$_{x,y}$}$ non-Dirac dispersion occurs at the high symmetry $\Gamma$ point with the fully spin-polarization. When the spin-orbit coupling is considered, the quadratic non-Dirac point opens the nontrivial topological gap of 2.70 - 159.10 meV. Remarkably, in the quantum anomalous Hall state both the chiral edge currents direction and the sign of Chern number can be tunable by doping. 4$\%$ compressive strain can induce structural phase transition but the nontrivial topological properties are maintained in the 2D BaX systems. Our studies will facilitate the experimental investigations on the spintronics applications.

\section*{Conflicts of interest}
There are no conflicts to declare.

\section*{Acknowledgements}
This work was carried out at Lvliang Cloud Computing Center of China, and the calculations were performed on TianHe-2.

\end{document}